\documentclass[twocolumn,showpacs,preprintnumbers,amsmath,amssymb,superscriptaddress]{revtex4}

\usepackage{graphicx}
\usepackage{bm}

\newcommand{\nc}{\newcommand}

\nc{\rnc}{\renewcommand}
\nc{\beq}{\begin{equation}}
\nc{\eeq}{\end{equation}} \nc{\beqa}{\begin{eqnarray}}
\nc{\eeqa}{\end{eqnarray}}

\begin{document}


\title[Supersymmetric dualities]
{\bf Supersymmetric dualities beyond the conformal window}

\author{V.~P.~Spiridonov}
\address{Bogoliubov Laboratory of Theoretical Physics,
JINR, Dubna, Moscow Region 141980, Russia; e-mail address:
spiridon@theor.jinr.ru}

\author{G.~S.~Vartanov}
\affiliation{Max-Planck-Institut f\"ur Gravitationsphysik, Albert-Einstein-Institut
14476 Golm, Germany; e-mail address: vartanov@aei.mpg.de}

\begin{abstract}
\medskip
Using the superconformal (SC) indices techniques, we construct Seiberg type
dualities for $\mathcal{N}=1$ supersymmetric field theories
outside the conformal windows. These theories are physically
distinguished by the presence of chiral superfields with small
or negative $R$-charges.
\end{abstract}


\maketitle

\section{Introduction}

Some of 4D ${\mathcal N}=1$ supersymmetric gauge field theories
are related by the Seiberg duality \cite{Seiberg}.
A full list of presently known dualities of such type for simple
gauge groups $G_c=SU(N), SP(2N), G_2$ is given in \cite{SV2}.
Remarkably, many of the listed dualities are {\em new}. Their
discovery is based on the interplay between superconformal (SC) indices
of \cite{Romelsberger1,Kinney,Romelsberger2} and the theory of
elliptic hypergeometric integrals formulated in \cite{S1,S2}
(see also \cite{S3}).

The $SU(2,2|1)$ space-time symmetry group is generated by $J_i, \overline{J}_i$
($SU(2)$ subgroups generators, or Lorentz rotations),
$P_\mu, Q_{\alpha},\overline{Q}_{\dot\alpha}$ (supertranslations),
$K_\mu, S_{\alpha},\overline{S}_{\dot\alpha}$ (special
superconformal transformations),
$H$ (dilations) and $R$ ($U(1)_R$-rotations).
For a distinguished pair of supercharges, say, $Q=\overline{Q}_{1 }$
and $Q^{\dag}=-{\overline S}_{1}$, one has
\begin{equation}
\{Q,Q^{\dag}\}= 2{\mathcal H},\quad \mathcal{H}=H-2\overline{J}_3-3R/2,
\label{susy}\end{equation}
and the SC index is defined by the matrix integral
\begin{eqnarray}\nonumber
&& I(p,q,f_k) = \int_{G_c} d \mu(g)\,
Tr \Big( (-1)^{\rm F}
p^{\mathcal{R}/2+J_3}q^{\mathcal{R}/2-J_3}
\\ && \makebox[1em]{} \times
e^{\sum_{a} g_aG^a} e^{\sum_{k}f_kF^k}e^{-\beta {\mathcal H}}\Big),
\quad \mathcal{R}= H-R/2,
\label{Ind}\end{eqnarray}
where $d \mu(g)$ is the $G_c$-invariant measure and ${\rm F}$ is the
fermion number operator. Operators $G^a$ and $F^k$
are the gauge and flavor group generators; $p,q,g_a,f_k,\beta$ are
group parameters (chemical potentials). The trace is taken over
the whole space of states, but, because the operators used in
(\ref{Ind}) preserve relation (\ref{susy}), only the zero modes
of the operator $\mathcal H$ contribute to the trace (hence, formally
there is no dependence on $\beta$).

The key idea of R\"omelsberger \cite{Romelsberger2} on the equality of
SC indices (\ref{Ind}) for the Seiberg dual theories
was realized first by Dolan and Osborn for a number of examples \cite{DO}.
These equalities are expressed in terms of the exact
computability of elliptic beta integrals discovered in \cite{S1} or
nontrivial symmetry transformations for higher order
elliptic hypergeometric functions on root systems \cite{S2,Rains}.

In addition to the description of new ${\mathcal N}=1$ dualities from known
identities for integrals, another important result of \cite{SV2}
consisted in the formulation of new mathematical conjectures
for integral identities following from known dualities. There are
also examples when both the dualities and corresponding relations for
integrals (indices) are new. The power of the theory of elliptic
hypergeometric integrals in application to the SC indices
techniques was demonstrated also in recent papers
by Gadde et al \cite{GPRR1,GPRR2}.

Here we focus on some physical consequences following from
the considerations of \cite{SV2}. Namely, we concentrate on implications
for the conformal windows introduced in \cite{Seiberg,IP}.
In the original Seiberg work \cite{Seiberg} it was shown that the
corresponding $G_c=SU(N)$ SQCD duality has distinguished
properties if the number of colors $N$
and the number of chiral superfields (flavors) $N_f$ satisfy
\beq\label{window_SU}
3N/2 < N_f < 3N.
\eeq
This conformal window guarantees that both dual theories have
asymptotic freedom and represent interacting SC theories at the
IR fixed points.
For $SP(2N)$ gauge groups with $N_f$ flavors the conformal window is \cite{IP}
\beq
3(N+1)/2 < N_f < 3 (N+1).
\label{window_SP}\eeq
After some time it started to be believed that these conformal windows
serve as the general necessary conditions for the existence of dualities
between interacting gauge theories. Our goal is to describe some multiple
dualities which do not fit this expectation.

Equality of SC indices of dual theories is a new
non-trivial indication on the validity of Seiberg dualities.
Earlier there were only the following justifying arguments \cite{Seiberg}.

1. The 't Hooft anomaly matching conditions. They were
conjectured in \cite{SV2} to be a consequence of the so-called
total ellipticity condition for the elliptic hypergeometric integrals \cite{S3}
describing SC indices.

2.  Matching reduction of the number of flavors $N_f\to N_f-1$.
Integrating out $k$-th flavor quarks by the mass term
$M_k^{\ k} Q_k \widetilde{Q}^k$ in the original theory results in
Higgsing the magnetic theory gauge group with a reduction of the additional
meson fields.  From the elliptic hypergeometric integrals point of view
this is realized by restricting in a special way a pair of parameters
($s_kt_k=pq$) which reduces the indices appropriately.

3. Matching of the moduli spaces and gauge invariant operators
in dual theories. Perhaps, this information is hidden in the
topological meaning of SC indices.

\section{$SU(N)$ gauge group}

{\bf A. $SU(2N)$ gauge group with $N_f=4$.} The starting electric
theory has $G_c=SU(2N)$ and the matter fields content $4f +
4\overline{f} + T_A +\overline{T}_A$, where $f$ and $T_A$  denote
the fundamental and absolutely antisymmetric tensor representations
of $G_c$ (the bar means conjugate representations). The flavor group
for $N>2$ is $SU(4) \times SU(4) \times U(1)_1 \times U(1)_2 \times
U(1)_B$. The SC index is given by the following integral \cite{SV2}
\begin{eqnarray}\label{SU2N_1}
&&I_E= \kappa_N \int_{\mathbb{T}^{2N-1}} \prod_{1 \leq i<j \leq 2N}
\frac{\Gamma(Uz_iz_j,Vz_i^{-1}z_j^{-1};p,q)}{\Gamma(z_i^{-1}z_j,z_iz_j^{-1};p,q)}
\nonumber\\  && \makebox[3em]{} \times \prod_{j=1}^{2N}
\prod_{k=1}^{4} \Gamma(s_kz_j,t_kz_j^{-1};p,q)\prod_{j=1}^{2N-1}
\frac{dz_j}{2\pi \textup{i} z_j},
\end{eqnarray}
where $\prod_{j=1}^{2N}z_j=1$, $\mathbb{T}$ is the unit circle with
positive orientation, $|U|, |V|, |s_k|, |t_k|<1$, and
$(UV)^{2N-2}\prod_{k=1}^4 s_kt_k = (pq)^2$. We use conventions
$\Gamma(a,b;p,q)\equiv\Gamma(a;p,q)\Gamma(b;p,q),$
$\Gamma(az^{\pm1};p,q)\equiv\Gamma(az;p,q)\Gamma(az^{-1};p,q)$, where
$$
\Gamma(z;p,q)= \prod_{i,j=0}^\infty
\frac{1-z^{-1}p^{i+1}q^{j+1}}{1-zp^iq^j}, \quad |p|, |q|<1,
$$
is the elliptic gamma function. Finally,
$$
\kappa_N = \frac{(p;p)_{\infty}^{2N-1} (q;q)_{\infty}^{2N-1}}{(2N)!}
$$
with $(a;q)_\infty=\prod_{k=0}^\infty(1-aq^k)$.
The parameters $U,$ $V,$ $s_k,$ $t_k$ are related to $f_k$ in (\ref{Ind})
and $z_j$ replace $g_a$.

In \cite{SV2} we described three magnetic duals for this model (one
of which was found earlier in \cite{C2}).
Equality of the corresponding SC indices is not proven yet, though
their $N_f=3$ simplifications do coincide, as follows from the
identities established in \cite{S2}. The dualities beyond the
conformal window  of interest emerge after some ``reduction"
of these theories. Namely, we restrict the parameters $U$ and
$V$ in (\ref{SU2N_1}) by the constraint $UV  =pq.$ Now
$\prod_{k=1}^4 s_kt_k = (pq)^{4-2N}$ and some of the parameters have
modulus bigger than 1. In this case it is necessary to use the
analytical continuation of integral (\ref{SU2N_1}) reached by
passing from $\mathbb T$ to a contour separating sequences of
integrand's poles converging to zero from their reciprocals. Due to
the inversion formula $\Gamma(z,pqz^{-1};p,q)=1$, the parameters $U$
and $V$ disappear completely from the electric SC index. As a
result, it becomes equal to the index of the theory without the
fields $T_A$ and $\overline{T}_A$ and global $U(1)_1\times U(1)_2$
symmetry, which coincides with the Seiberg
electric theory with $N_f=4$ \cite{Seiberg}.
The type I $A_N$-elliptic beta integral evaluation \cite{S3} shows that
for $N>1$ the reduced SC index is equal to zero.

The dual magnetic theories are reduced in a similar way. We substitute
into the magnetic indices described in \cite{SV2} $U=\sqrt{pq}x,\
V=\sqrt{pq}x^{-1}$, where $x$ is the chemical potential
of the $U(1)_1$-group, and interpret them as the indices
of reduced theories. The fields content and some of the $R$-charges
of the resulting theories differ from the original ones.
As a result, we find the following set of dualities.
First magnetic theory is described in Tab. 1.
\begin{center}
\begin{tabular}{|c|c|c|c|c|c|c|}
  \hline
   & $SU(2N)$ & $SU(4)$ & $SU(4)$ & $U(1)_1$ & $U(1)_B$ & $U(1)_R$ \\  \hline
  $q$ & $f$ & $f$ & 1 & 0 & -1 & $-\frac 12(N-2)$ \\
  $\widetilde{q}$ & $\overline{f}$ & 1 & $f$ & 0 & 1 & $-\frac 12(N-2)$
  \\
  $H_m$ & 1 & $T_A$ & 1 & -1 & 2 & $2m-N+3$ \\
  $G$ & 1 & $T_A$ & 1 & $N-1$ & 2 & 1 \\
  $\widetilde{H}_m$ & 1 & 1 & $T_A$ & 1 & -2 & $2m-N+3$ \\
  $\widetilde{G}$ & 1 & 1 & $T_A$ & $1-N$ & -2 & 1 \\
\hline
\end{tabular}
\\ \small{Tab. 1. First $SU(2N)$ dual theory, where $m=0,\ldots,N-2.$}
\end{center}
\begin{center}
\begin{tabular}{|c|c|c|c|c|c|}
  \hline
   & $SU(2N)$ & $SU(4)$ & $SU(4)$ & $U(1)_B$ &  $U(1)_R$ \\  \hline
  $q$ & $f$ & $\overline{f}$ & 1 & 1 & $-\frac 12(N-2)$ \\
  $\widetilde{q}$ & $\overline{f}$ & 1 & $\overline{f}$ & -1 & $-\frac 12(N-2)$
  \\
  $M_k$ & 1 & $f$ & $f$ & 0 & $2k-N+2$ \\
\hline
\end{tabular}
\\ \small{Tab. 2. Second $SU(2N)$ dual theory, where $k=0,\ldots,N-1$.}
\end{center}
\begin{center}
\begin{tabular}{|c|c|c|c|c|c|c|c|}
  \hline
   & $SU(2N)$ & $SU(4)$ & $SU(4)$ & $U(1)_1$ & $U(1)_B$ & $U(1)_R$ \\  \hline
  $q$ & $f$ & $\overline{f}$ & 1 & 0 & -1 & $-\frac 12(N-2)$ \\
  $\widetilde{q}$ & $\overline{f}$ & 1 & $\overline{f}$ & 0 & 1 & $-\frac 12(N-2)$
  \\
  $M_k$ & 1 & $f$ & $f$ & 0 & 0 & $2k-N+2$ \\
  $H_m$ & 1 & $T_A$ & 1 & -1 & 2 & $2m-N+3$ \\
  $G$ & 1 & $T_A$ & 1 & $N-1$ & 2 & 1 \\
  $\widetilde{H}_m$ & 1 & 1 & $T_A$ & 1 & -2 & $2m-N+3$ \\
  $\widetilde{G}$ & 1 & 1 & $T_A$ & $1-N$ & -2 & 1 \\
\hline
\end{tabular}
\\ \small{Tab. 3. Third $SU(2N)$ dual theory,
where $k=0,\ldots,N-1$ and $m=0,\ldots,N-2.$}
\end{center}
In all our tables the first column contains symbols of the fields and
the second---the gauge group representations. For $U(1)$ groups
we give corresponding
hypercharges. We skip also the vector superfield and its duals
(adjoint representations of $G_c$ and singlets of the flavor groups).

The global symmetry and field content of the second magnetic
theory is the same as in Seiberg's dual theory with $N_f=4$ (see
Tab.  2), but the gauge group is now $SU(2N)$ instead of
$SU(N_f-2N)$. The most complicated is the third magnetic theory
(see Tab. 3). SC indices of all these magnetic duals vanish for $N>1$,
which coincides with the electric index.
For $N=1$ we come to the family of dualities
considered in detail in \cite{SV1}.

{\bf B. $SU(N)$ gauge group with $N_f=N+2$.}
The electric part of the next set of dualities coincides with the
Seiberg theory for $N_f=N+2$ and arbitrary $N$. Its canonical
magnetic dual has $G_c=SU(2)$, and it is IR free for $N>4$ \cite{Seiberg}.

Our new magnetic dual theories have $G_c=SU(N)$ and the
flavor symmetry group $SU(K)\times SU(M) \times
U(1)_1 \times SU(K) \times SU(M) \times U(1)_2 \times U(1)_B$, where
$M=N+2-K$ and $K=1,\dots,N+1$. For the field content see Tab. 4.

These dualities were derived in \cite{SV2} (for $N=2$, see \cite{SV1})
from the equality of SC indices of the corresponding theories, which
follows from the identities established by Rains \cite{Rains}
(for $K=1$, see \cite{S2}).
Here we just stress that they lie outside the conformal window
(\ref{window_SU}) for $N>3$, since the left-hand side inequality is
violated. Surprisingly, for $N=3$
we obtain a new duality lying {\it inside} the conformal window.

\begin{widetext}
\begin{center}
\begin{tabular}{|c|c|c|c|c|c|c|c|c|c|}
  \hline
   & $SU(N)$ & $SU(K)$ & $SU(M)$ & $U(1)_1$ & $SU(K)$ & $SU(M)$ & $U(1)_2$ & $U(1)_B$ & $U(1)_R$ \\  \hline
  $q_1$ & $\overline{f}$ & $f$ & 1 & $\frac{K(K-2)}{N}-K+M$ & 1 & 1 & $\frac{MK}{N}$ & $1-M$ & $\frac{2}{N+2}$ \\
  $q_2$ & $f$ & 1 & $f$ & $-\frac{K(K-2)}{N}$ & 1 & 1 & $\frac{-MK}{N}$ & $1-K$ & $\frac{2}{N+2}$ \\
  $q_3$ & $f$ & 1 & 1 & $\frac{MK}{N}$ & $f$ & 1 & $\frac{K(K-2)}{N}-K+M$ & $M-1$ & $\frac{2}{N+2}$ \\
  $q_4$ & $\overline{f}$ & 1 & 1 & $-\frac{MK}{N}$ & 1 & $f$ & $-\frac{K(K-2)}{N}$ & $K-1$ & $\frac{2}{N+2}$ \\
  $X_1$ & 1 & $f$ & 1 & $M$ & 1 & $f$ & $-K$ & 0 & $\frac{4}{N+2}$ \\
  $X_2$ & 1 & 1 & $f$ & $-K$ & $f$ & 1 & $M$ & 0 & $\frac{4}{N+2}$ \\
  $Y_1$ & 1 & $\overline{f}$ & $\overline{f}$ & $K-M$ & 1 & 1 & 0 & $N$ & $\frac{2N}{N+2}$ \\
  $Y_2$ & 1 & 1 & 1 & 0 & $\overline{f}$ & $\overline{f}$ & $K-M$ & $-N$ & $\frac{2N}{N+2}$ \\
\hline
\end{tabular}
\\ \small{Tab. 4. $SU(N)$ magnetic theories with $N+2$ flavors.}
\end{center}
\end{widetext}

\section{$SP(2N)$ gauge group}

We describe now dualities lying outside the conformal window (\ref{window_SP}).
The starting electric theory has $G_c=SP(2N)$ and the matter fields $8f+T_A$.
As shown in \cite{SV1}, this theory has many dual partners
(one of which was found earlier in \cite{C1}). The electric
SC index has the form
\begin{eqnarray}\label{SP2N1}\nonumber
    && I_E = \kappa_N
\Gamma(t;p,q)^{N-1} \int_{{\mathbb T}^N}\prod_{1 \leq i < j \leq N}
\frac{\Gamma(t z_i^{\pm 1} z_j^{\pm 1};p,q)} {\Gamma(z_i^{\pm 1}
z_j^{\pm 1};p,q) }
    \\     &&   \makebox[4em]{} \times
\prod_{j=1}^N \frac{\prod_{k=1}^8  \Gamma(t_k z_j^{\pm 1};p,q)}
{\Gamma(z_j^{\pm 2};p,q)} \frac{d z_j}{2 \pi \textup{i} z_j},
\end{eqnarray}
where $|t|, |t_k|<1$, $t^{2N-2} \prod_{k=1}^8 t_k = (pq)^2$,
and
$$
 \kappa_N \ = \ \frac{(p;p)_{\infty}^N (q;q)_{\infty}^N }{2^N N!}.
$$
This integral has nice symmetry transformations described by the Weyl group of
the exceptional root system $E_7$ \cite{Rains} (for $N=1$, see \cite{S2}).

Now we restrict the $t$-parameter value to $t=\sqrt{pq}$ and
analytically continue function (\ref{SP2N1}) by replacing $\mathbb
T$ to a contour separating geometric sequences of integrand's poles
converging to zero from their reciprocals. This leads to the
``decoupling" of the $T_A$-field from the electric theory, so that the
same index is generated by the model with $8$ quarks in fundamental
representations of $G_c$ and flavor group $SU(8)$
with the $R$-charge equal to $(3-N)/4$.

To obtain the dual description, we set $t=\sqrt{pq}$ in the magnetic
SC indices \cite{SV1} and interpret the resulting
integrals as coming from different dual theories, similar to the
$SU(2N)$ case described above. The field content of first
magnetic theory is given in Tab. 5 (note the change of the flavor
group).
\begin{center}
\begin{tabular}{|c|c|c|c|c|c|}
  \hline
                & $SP(2N)$            & $SU(4)$    & $SU(4)$    & $U(1)_B$ & $U(1)_R$
\\  \hline
  $q$             & $f$                & $f$            & 1            & $-1$   &   $-\frac{N-3}{4}$
  \\
$\widetilde{q}$& $f$   & 1            &$f$&  $1$   &
$-\frac{N-3}{4}$
\\
  $M_J  $             & 1         & $T_A$            &    1       & $2$     &  $J-\frac{N-3}{2}$
  \\
  $\widetilde{M}_J $    & 1      & 1            &    $T_A$       & $-2$    &
  $J-\frac{N-3}{2}$ \\
\hline
\end{tabular}
\\ \small{Tab. 5. First $SP(2N)$ dual theory, where $J=0,\ldots,N-1$.}
\end{center}
Second and third magnetic theories are described in Tabs. 6 and 7.
The third theory was found in \cite{IP}, its flavor group coincides
with the electric one.
Note that SC indices of all four dual theories are equal to zero for $N>2$,
as follows from vanishing of the type I $BC_N$-elliptic beta integral
for $N_f<N+2$ \cite{S3}.

\begin{center}
\begin{tabular}{|c|c|c|c|c|c|}
  \hline
                & $SP(2N)$            & $SU(4)$    & $SU(4)$    & $U(1)_B$ & $U(1)_R$            \\  \hline
  $q$           & $f$                & $\overline{f}$   & 1   & $1$     &   $-\frac{N-3}{4}$  \\
$\widetilde{q}$& $f$   & 1            &$\overline{f}$&
$-1$   &   $-\frac{N-3}{4}$ \\
  $M_J $          & 1              & $f$            &    $f$         & 0    & $J-\frac{N-3}{2}$ \\
\hline
\end{tabular}
\\ \small{Tab. 6. Second $SP(2N)$ dual theory with $8$
flavors.}
\end{center}
\begin{center}
\begin{tabular}{|c|c|c|c|}
  \hline
   & $SP(2N)$ & $SU(8)$ & $U(1)_R$ \\  \hline
  $q$ & $f$ & $\overline{f}$ & $-\frac{N-3}{4}$ \\
  $M_J $ & 1 & $T_A$ & $J-\frac{N-3}{2}$ \\
\hline
\end{tabular}
\\ \small{Tab. 7. Third $SP(2N)$  dual theory with $8$
flavors.}
\end{center}

\section{Conclusion}

For all new dualities described in this paper we have checked
validity of the 't Hooft anomaly matching conditions.
As mentioned already,  they pass also the new duality test by
having equal SC indices.

The first and third magnetic duals of Sect. IIA are rather
unusual---they have the additional $U(1)_1$-group, which does not
interact with the quarks and whose anomalies vanish.
Vanishing of the indices of theories in Sect. IIA for $N>1$ and Sect. III for
$N>2$ indicates that these models are similar to
the Seiberg $SU(N)$ electric theory with $N_f\leq N$ (e.g.,
they may have problems with the ground state).
The $G_c=SP(4)$ case of Sect. III is interesting as well.
Corresponding electric theory is confining \cite{IP}, which
means that all our other dual theories (which were missed in \cite{IP})
also confine. Their SC indices obey $W(E_7)$-symmetry and can be
evaluated explicitly \cite{S3}, in difference from the $SP(2)$-group
case \cite{SV1}.

As to the new dualities of Sect. IIB, their  origin is quite simple. The
$f$ and $\overline{f}$ representations of the dual $SU(2)$ gauge group are
equivalent, and the corresponding flavor group gets enlarged from
$SU(N_F)\times SU(N_F)\times U(1)_B$ to $SU(2N_F)$. Permuting
corresponding character variables in an arbitrary way, one can
construct ``duals of duals" with $G_c=SU(N)$ in many different ways.
Although this is a rather evident possibility, it was missed in
the previous discussions of the Seiberg duality.
We remark also that all the models described in our tables
are asymptotically free and define interacting conformal
field theories at the IR fixed point.

We conclude that the notion of conformal windows should be used
with care --- it is applicable only to particular types of
dualities. Our results raise a natural question on classification
of all 4D theories dual to the original Seiberg
``minimal" electric SQCD. It is necessary to analyze various IR
physics implications following from the described dualities.
In particular, this concerns the structure of superpotentials (see, e.g.,
\cite{Kh}). It would be interesting to
understand which properties of the SC indices
are responsible for the description of moduli spaces and
natural choices of the superpotentials. Equalities of
indices of dual theories remain valid away from the IR fixed points.
This and other mathematical properties of SC indices raise the problem
of establishing all physical information hidden in them.

For SC field theories (e.g., ${\mathcal N}=1$ theories
at the IR fixed points), the dimension of the
scalar component of a gauge invariant chiral
superfield is related to its $R$-charge as
$\Delta =3R/2$. For the meson field $M=Q\widetilde{Q}$ with $G_c=SU(N)$
the dimension is $\Delta[M] = \Delta[Q] + \Delta[\widetilde{Q}] =
3R =3 (1-N/N_f)$.  The conventional  SC algebra
wisdom on unitarity demands that $\Delta[M] \geq 1$, or $N_f\geq 3N/2$,
which is clearly broken in our theories for $N>4$. Therefore one has to
find physical ways out of this obstacle
either by modifying the IR dynamics or by other means.
The theories of Sect. IIB are unitary for $N=2,3,4$;
the new $SU(3)$ duality satisfies thus all physical requirements
and deserves further detailed investigation.

We thank L. Alvarez-Gaum\'e, Z. Komargodski,
J.~M. Mal\-da\-ce\-na, M.~A. Shif\-man, M. Sudano, and B. Wecht
for valuable discussions. V.S. was partially supported by
RFBR grant no. 09-01-00271,  CERN TH (Geneva), and MPIM (Bonn).
G.V. would like to thank TPI (Minnesota),
IAS (Princeton), and CNYITP (Stony Brook) for hospitality during
his visit in March 2010.


\begin{thebibliography}{99}

\bibitem{Seiberg} N. Seiberg, Nucl. Phys. {\bf B435} (1995), 129.

\bibitem{SV2}  V.~P.~Spiridonov and G.~S.~Vartanov,
\textit{Elliptic hypergeometry of supersymmetric dualities},
arXiv: 0910.5944.

\bibitem{Romelsberger1} C. R\"omelsberger, Nucl. Phys. {\bf B747} (2006), 329.

\bibitem{Kinney} J. Kinney, J. M. Maldacena, S. Minwalla and S.~Raju,
Commun. Math. Phys. {\bf 275} (2007), 209.

\bibitem{Romelsberger2} C. R\"omelsberger,
\textit{Calculating the Superconformal Index and Seiberg Duality},
arXiv: 0707.3702.

\bibitem{S1} V. P. Spiridonov, Russian Math. Surveys {\bf 56} (1) (2001), 185.

\bibitem{S2} V. P. Spiridonov,
Algebra i Analiz {\bf 15} (6) (2003), 161, math.CA/0303205.

\bibitem{S3} V. P. Spiridonov,
Russian Math. Surveys {\bf 63} (3) (2008), 405, arXiv: 0805.3135.

\bibitem{DO} F. A. Dolan and H. Osborn, Nucl. Phys. {\bf B818} (2009), 137.

\bibitem{Rains} E. M. Rains, Ann. of Math. {\bf 171} (2010), 169.

\bibitem{GPRR1} A.~Gadde, E.~Pomoni, L.~Rastelli and S.~S.~Razamat,
JHEP {\bf 03} (2010) 032.

\bibitem{GPRR2} A.~Gadde, L.~Rastelli, S.~S.~Razamat and W.~Yan,
\textit{The Superconformal Index of the $E_6$ SCFT}, arXiv: 1003.4244.

\bibitem{IP} K. A. Intriligator and P. Pouliot,
Phys. Lett. {\bf B353} (1995), 471.

\bibitem{C2} C. Cs\'aki, M. Schmaltz, W. Skiba and J. Terning,
Phys. Rev. {\bf D56} (1997), 1228.

\bibitem{SV1} V. P. Spiridonov and G. S. Vartanov,
Nucl. Phys. {\bf B824} (2010), 192.

\bibitem{C1} C. Cs\'aki, W. Skiba and M. Schmaltz,
Nucl. Phys. {\bf B487} (1997), 128.

\bibitem{Kh} A. Khmelnitsky, JHEP {\bf 03} (2010), 065.

\end{thebibliography}
\end{document}